\documentclass[12pt]{article}
\usepackage{scicite}
\usepackage{times}
\usepackage{graphicx}
\usepackage{amsmath}
\usepackage[usenames,dvipsnames]{color}
\usepackage[export]{adjustbox}
\usepackage{amsmath}
\usepackage{amssymb}

\newenvironment{sciabstract}{
\begin{quote} \bf}
{\end{quote}}

\newcounter{lastnote}

\title{Entanglement transition in random rod packings}

\author{Yeonsu Jung$^1$, Thomas Plumb-Reyes$^1$,  Hao-Yu Greg Lin$^2$, \\ and L. Mahadevan$^{1,3,4\ast}$
\\
\footnotesize{$^1 $John A. Paulson School of Engineering and Applied Sciences,}\\
\footnotesize{$^2$Center for Nanoscale Systems,}\\
\footnotesize{$^3$Department of Physics,}\\
\footnotesize{$^4$Department of Organismic and Evolutionary Biology, Harvard University, Cambridge, MA 02138, USA}\\
\footnotesize{$^\ast$To whom correspondence should be addressed; E-mail:  lmahadev@g.harvard.edu}
}

\date{}
\begin{document}

\baselineskip16pt
\maketitle

\begin{sciabstract}
Random packings of stiff rods are self-supporting mechanical structures stabilized by long range interactions induced by contacts. To understand the geometrical and topological complexity of the packings, we first deploy X-ray computerized tomography to unveil the structure of the packing. This allows us to directly visualize the spatial variations in density, orientational order and the entanglement, a mesoscopic field that we define in terms of a local average crossing number, a measure of the topological complexity of the packing. We find that increasing the aspect ratio of the constituent rods in a packing leads to a proliferation of regions of strong entanglement that eventually percolate through the system, and correlated with a sharp transition in the mechanical stability of the packing. To corroborate our experimental findings, we use numerical simulations of contacting elastic rods and characterize their stability to static and dynamic loadings. Our experiments and computations lead us to an entanglement phase diagram which we also populate using published experimental data from pneumatically tangled filaments, worm blobs, and bird nests along with additional numerical simulations using these data sets. Together, these show the regimes associated with mechanically stable entanglement as a function of the statistics of the packings and loading, with lessons for a range of systems from reconfigurable architectures and textiles to active morphable filamentous assemblies.

\end{sciabstract}

\newpage

\section*{Introduction}\label{sec1}

The structure and response of polydisperse granular systems consisting of \textit{localized} particles, such as spheres and ellipsoids, can be characterized via their number density and positional disorder. In these systems, at sufficiently large densities, there is a transition to a disordered jammed packing with emergent mechanical properties that are controlled by a topological order parameter, the average coordination number \cite{ohern_random_2002,ohern_jamming_2003,song_phase_2008,Liu2010,liu_2014,Baule2018}. However, experimentally, spheroidal or ellipsoidal packings need to be confined by boundaries to see these effects; the packings are mechanically unstable in the absence of confining constraints. This is in sharp contrast with the well known behavior of disordered packings of long, thin rods seen in the familiar example of a bird nest or a randomly packed felt that forms solid-like states by jamming at packing fraction that are much smaller than those associated with {and ordered packings seen in any textile} \cite{Ekman2014,Philipse1996,bi_jamming_2011,weiner_mechanics_2020,Bhosale2022}.

The \textit{extended} nature of individual filaments that can entangle naturally implies the presence of topological constraints associated with contact. While these have been well studied in thermalized systems exemplified by polymer melts \cite{Onsager1949,Edwards1967,Doi1978}, their athermal macroscopic analogs seen in textiles \cite{hearle_physical_2008}, felts \cite{kabla_nonlinear_2007} etc. are much less understood, although recent work has started to highlight the importance of topological constraints\cite{buck_spectrum_2012,panagiotou_knot_2020,patil_topological_2020,glover_measuring_2024} in rod packings of large aspect ratio filaments \cite{trepanier_column_2010,franklin_geometric_2012,heussinger_collapse_2023} or curved filaments\cite{Gravish2012,Zhao2016,wang_structured_2021,karapiperis_stress_2022} with applications to soft robotics \cite{becker_active_2022} and active systems \cite{raviv2014active,patil2022ultrafast,bozdag_novo_2023}. The most basic system of this kind is a dense packing of isotropic rods where quenched disorder in orientation and contact prevents the system from being easily dismantled.

Spherical or spheroidal particles which have a single length scale, so that interactions are confined to the scale of the particle diameter. In contrast,  elongated particles have a diameter that is much smaller than their length, and their collective behavior is dominated by  non-local interactions since an individual filament can weave together many otherwise unconnected filaments.  This fundamental difference in interaction range suggests that traditional mean-field approaches, which are often used to describe collective behavior in systems of spherical particles, may not be applicable to systems of rod-like particles.

Here, we use a combination of X-ray tomography experiments and numerical simulations to study a dense packing of stiff rods of large aspect ratio. Our packings are made of steel rods with varying length-to-diameter ratios, i.e. aspect ratios {$\alpha \in [30,200]$}, (see Tab.~S4), imaged using X-ray CT and processed to accurately locate the centerline of the rods and their contacts (see \textbf{Materials and Methods} \ref{sec:image-segmentation}). This allows us to quantify the local volume fraction, orientational order, contact density, and various measures of entanglement and correlate them with the mechanical stability of the packing. To understand our results, we also employ numerical simulations of frictionally interacting filaments to support our experimental findings, enabling the exploration of phase space regions beyond our experimental reach. We then compare our experimental and computational results  with published experiments across a range of other entangled systems, which we summarize in terms of a phase diagram that delineates various entanglement regimes based on packing statistics and mechanical loading conditions.

\section*{Results}\label{sec2}

\subsection*{Stability of packings}

\begin{figure*}
  \centering
  \makebox[\textwidth][c]{\includegraphics[scale = 1]{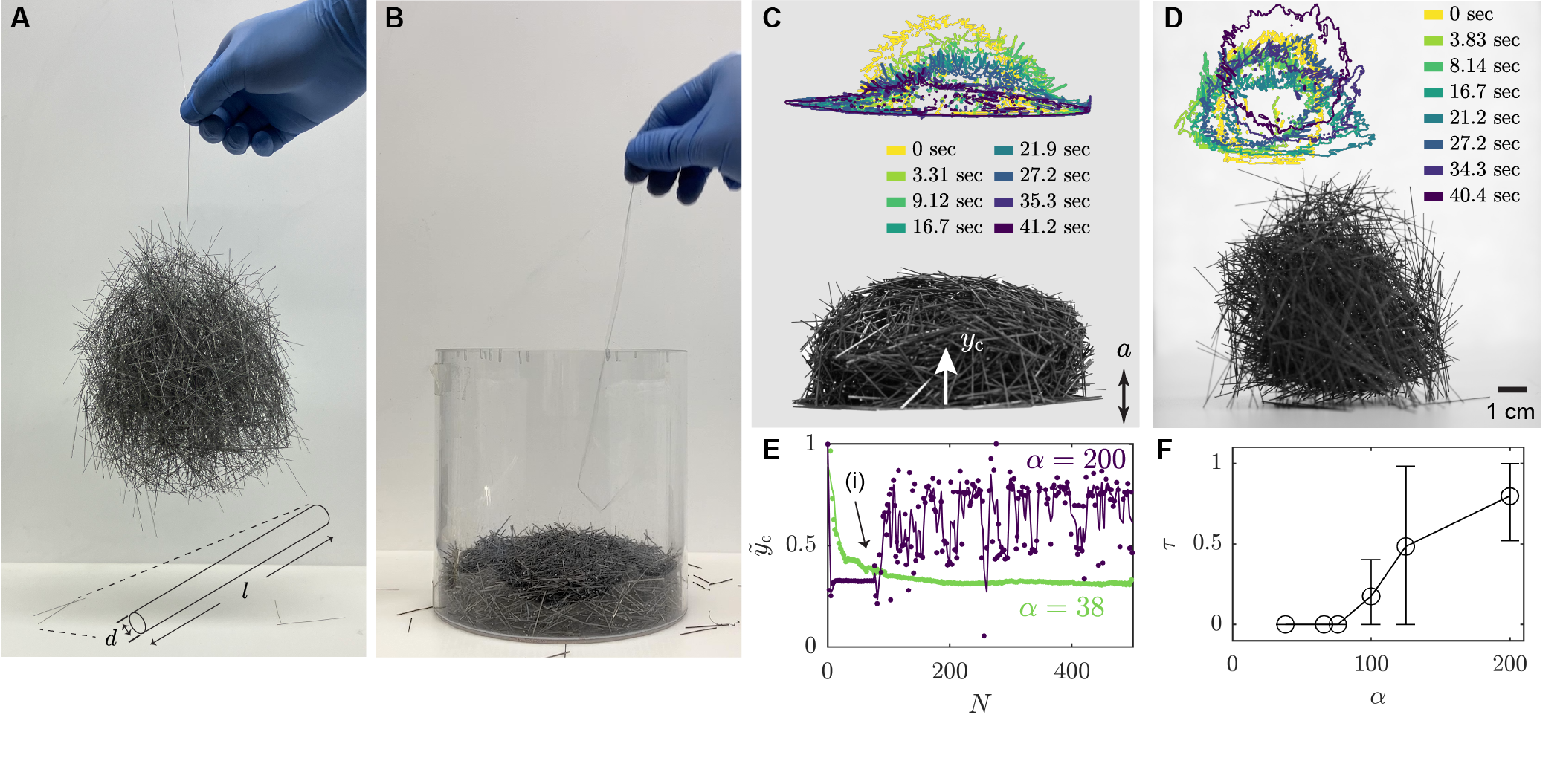}}
  \caption{\footnotesize{{\bf Formation and stability of entangled rods.} (A-B) Spontaneous formation of entanglement by sieving of rods (See Video S1 for details). Entangled rods of high aspect ratio ($\alpha = 200$) can support their own weight (A), while low aspect ratio rods ($\alpha = 38$) cannot (B). (C,D) Vertical excitation with a fixed frequency (5 Hz) and amplitude (25 mm) of rod packing with aspect ratios, $\alpha = 38$ (C) and $\alpha = 333$ (D). Inset shows that the large aspect ratio rod packing survives, but the small aspect ratio packing breaks down through contours of rod packings at each time instance. (E) Normalized centroid height of rods, $\tilde{y}_\mathrm{c} = y_\mathrm{c}(t)/y_\mathrm{c}(0)$, as a function of number of excitation cycles, $N = t \times f$ with time $t$ and excitation frequency $f$, of packings with two different aspect ratios 38 and 200. Formation (i) of an entangled unit during the excitation are denoted near the onset of them for $\alpha = 125$. (F) The ratio of entangled period to the observation time, $\tau = t_\mathrm{entangled}/T$, ($T = 100$ sec). $\tau$ serves as a measure of the persistence of the entangled phase under excitation. See Fig.~S1 which plots the raw data used to calculate $\tau$.}}
  \label{fig:Intro}
\end{figure*}

We prepared random packings of rods by pouring them gently into a cylindrical container through a mesh. Figure~\ref{fig:Intro}A and Video S1 show that such a packing of rods (diameter $d$, length $l$, aspect ratio $\alpha=l/d$), with $\alpha = 200$ can resist its own weight while a packing of rods with $\alpha = 38$ leads to an untangled pile that cannot sustain its own weight. We note that the presence of gravity breaks vertical symmetry, i.e. rods are likely to be heterogeneously arranged in this direction, confirmed by our later analysis of the packing geometry. 

To understand the stability of these packings, we subject them to harmonic excitation. In insets of Fig.~\ref{fig:Intro}C and D, we see that a high aspect-ratio rod packing ($\alpha = 200$) forms a dynamically stable clump, while a low aspect-ratio rod packing ($\alpha = 38$) collapses (for frequency, $f = 5$ Hz, and amplitude, $A = 25$ mm, so that the dimensionless maximum acceleration  {$a/g \equiv 4 \pi^2Af^2/g = 2.5$} where $g = 9.81$ m/s$^2$, although this behavior persists for a range of frequencies and amplitudes).

A method to quantify the stability of the entangled packings is afforded by the height of the clump as a function of time. In Fig.~\ref{fig:Intro}E, we see that the mechanical excitation can either destroy cohesion (for packings with $\alpha \le 100$), or enhance it (for packings with $\alpha \ge 100$). The formation of entangled units during excitation is likely due to the frequency-dependent nature of the transition. We observe that near the critical frequency, a small change in the frequency can lead to a large change in the outcome, i.e. the packing can switch from a stable mechanical collective to a collapsed configuration (see Video S2).

To compare the dynamic stability of rod packings as a function of the aspect ratio $\alpha$, we prepare samples based on the assumption that strongly excited rods in a closed container will have the isotropic orientation statistics and the same coordination number ($Z \approx 6$) \cite{Philipse1996,blouwolff_coordination_2006}, where the system volume fraction (the ratio of the rods' volume to the system volume) follows (see \textbf{Materials and Methods} \ref{sec:packing-preparation}):
\begin{equation}
    \Phi \sim Z \alpha.
    \label{eq:volume_fraction}
\end{equation}
Consequently, the number of rods introduced into the container depends on its aspect ratio (more specifically, $N \sim Z\frac{V}{d l^2}$ or $\rho \sim \frac{Z}{dl^2}$ with $\rho$  being number density, see \textbf{Materials and Methods} \ref{sec:packing-preparation}).

The response to harmonic excitation of these packings allows us to define a scaled time $\tau = t_\mathrm{entangled}/100~\mathrm{sec}$, the ratio of the duration for which the entangled phase persists relative to the observation time. In Fig.~\ref{fig:Intro}E, we show that $\tau \approx 0$ for $\alpha \ll 100$ but increases as $\alpha \ge 100$, a dynamical signature of an entanglement transition.

\subsection*{Mesoscopic measures of volume fraction, orientational order, average contact number and entanglement}

\begin{figure*}[h!]
  \centering    
  \makebox[\textwidth][c]{\includegraphics[scale = 1]{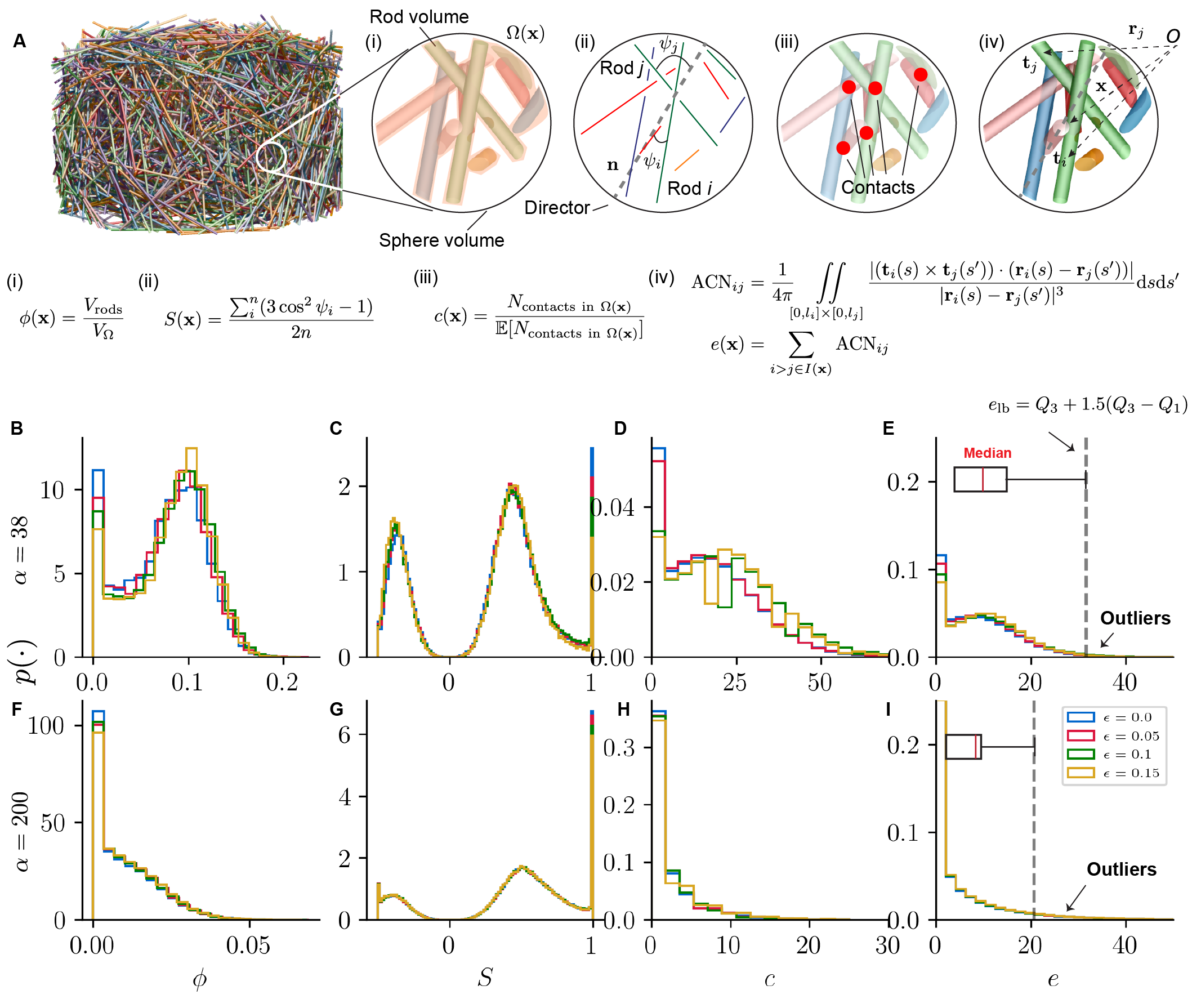}}    
  \caption{\footnotesize{{\bf Experimental characterization of geometry and topology of a dense rod packing.} (A) Reconstructed 3D rendering of X-ray CT scan of a local domain. For each rod $i$, we identify the centerline  $\mathbf{r}_i(s_i)$ and unit tangent vector $\mathbf{t}_i = \dot{\mathbf{r}}_i/|\dot{\mathbf{r}_i}|$. Volume fraction ($\phi$), orientational order ($S$), contact density ($c$), and entanglement ($e$) fields can be calculated using the Eqs.~S1-S5 where A(i) - A(iv) show corresponding graphical description of those fields. (B-I) Statistics of local volume fraction $\phi$ (B and F), orientational order $S$ (C and G), contact density, $c$ (D and H), and entanglement $e$ in Eqs.~S1-S5 (E and I) fields of the rod packing when $\epsilon = 0$ and $\epsilon = 0.15$. The lowerbounds, $e_\mathrm{lb}$, for outliers in $p(e)$ for $\alpha = 38$ and $100$ are shown in (E) and (I) where $e_\mathrm{lb}$ is the $Q_3 + 1.5(Q_3 - Q_1)$ with $Q_1$ and $Q_3$ being the lower and upper quartiles of the entanglement distributions. Inset: Box plots indicating $Q_1$, $Q_3$, $e_\mathrm{lb}$, mean, and median of $p(e)$.}}
  \label{fig:Fields}
\end{figure*}

X-ray CT and image analysis allow us to characterize the volume fraction $\phi$, orientational order $S$, contact density $c$, and entanglement $e$ (see Fig.~\ref{fig:Fields}A) as a function of location for different packing and associated conditions. For high aspect packings which possess two well-separated length scales, rod diameter $d$ and length $l$, a natural mesoscopic scale is the geometric mean, $R = (d l)^{1/2}$, especially for the purpose of comparing among packings of different aspect ratio (see SI for discussion of this scale).
The mesoscopic neighborhood $\Omega(\mathbf{x}) = \Omega_R(\mathbf{x})$ is a sphere centered around $\mathbf{x}$ with $R = (dl)^{1/2}$, with multiple rods within $\Omega(\mathbf{x})$ which we can locally label with the index $i \in I(\mathbf{x}) = \{1,2, \cdots, n\}$ denoting the local labels of rods in $\Omega(\mathbf{x})$. The number of rods $ n_R(\mathbf{x})$ in $\Omega(\mathbf{x})$  depends on the choice of $R$ but becomes independent of aspect ratio for isotropic packing because $\rho \sim Z/dl^2$ and $n \sim (Z/dl^2)(dl) (l) \sim Z.$ (see \textbf{Materials and Methods} \ref{sec:packing-preparation} for a derivation).

Given the clipped lengths of each rod $i$ denoted by $l_i < R$,  in $\Omega(\mathbf{x})$, the solid volume occupied by rods is  $V_\mathrm{rods} = \sum_i^n \pi d^2 l_i$. Then, the volume fraction of rods is given by
\begin{equation}
    \phi(\mathbf{x}) = \frac{V_\mathrm{rods}}{V_{\Omega}}
\end{equation}
with $V_{\Omega}=4\pi R^3/3$ is the volume of bounding sphere.

We can also define the scalar orientational order within $\Omega$ as,
\begin{equation}
    S(\mathbf{x}) = \frac{\sum_i^n (3\cos^2 \psi_i - 1)}{2 n}
\end{equation}  where $\psi_i$ is the angle between $i$th rod and the average orientation within $\Omega$;  completely aligned configuration will yield $S = 1$, while an isotropic one will yield $S = 0$.

We define the contact density $c$ in a local neighborhood within $\Omega$ as the number of contacts divided by the expected number of contacts in that volume, or
\begin{equation}
    c(\mathbf{x}) = \frac{N_\mathrm{contacts~in~\Omega(\mathbf{x})}}{\mathbb{E} [N_\mathrm{contacts~in~\Omega(\mathbf{x})}]}
\end{equation}
where $\mathbb{E} [N_\mathbf{contacts~in~\Omega(\mathbf{x})}]$ = $\rho Z V_\Omega/2$. The factor of half considers the fact that one contact count is shared by two rods, or $N_\mathrm{total~contacts} = NZ/2$.

Finally, we define the entanglement $e$ which quantifies how rods are intertwined within a local neighborhood by summing the average crossing number (ACN) between all possible pairs, inspired by the fact that it has been used to describe the complexity of filamentous arrangements \cite{buck_spectrum_2012,becker_active_2022}. A 3D configuration of a pair of spatial curves can be projected onto an arbitrary plane, revealing possible crossings in each projection. The ACN is the average number of crossings across all possible projections, so that given two rods labeled $i$ and $j$, the ACN is defined as \cite{buck_spectrum_2012}
\begin{equation}
    \mathrm{ACN}_{ij} = \frac{1}{4\pi} \iint\limits_{[0,l_i]\times[0,l_j]} \frac{|(\mathbf{t}_i(s) \times \mathbf{t}_j(s'))\cdot(\mathbf{r}_i(s) - \mathbf{r}_j(s'))|}{|\mathbf{r}_i(s) - \mathbf{r}_j(s')|^3} \mathrm{d}s \mathrm{d}s'
\end{equation}
where $\mathbf{r}_i(s): [0,l_i] \rightarrow \mathbb{R}^3$ and $\mathbf{t}_i(s) = \mathbf{r}'_i(s): [0,l_i] \rightarrow \mathbb{S}^2$ are the centerline and unit tangent vector of rod $i$, respectively, with the arc length parameter $s$.   To calculate $e$, we first identify the edges of each rod within the neighborhood. Then, we compute the average crossing number for each pair of edges and sum these values across all possible pairs, or
\begin{equation}
    e(\mathbf{x}) = \sum_{i>j} \mathrm{ACN}_{ij}
\end{equation}
where the labels, $i,j \in I(\mathbf{x})$, denote rods in $\Omega(\mathbf{x})$.

\begin{figure*}[h!]
  \centering    
  \makebox[\textwidth][c]{\includegraphics[scale = 0.70]{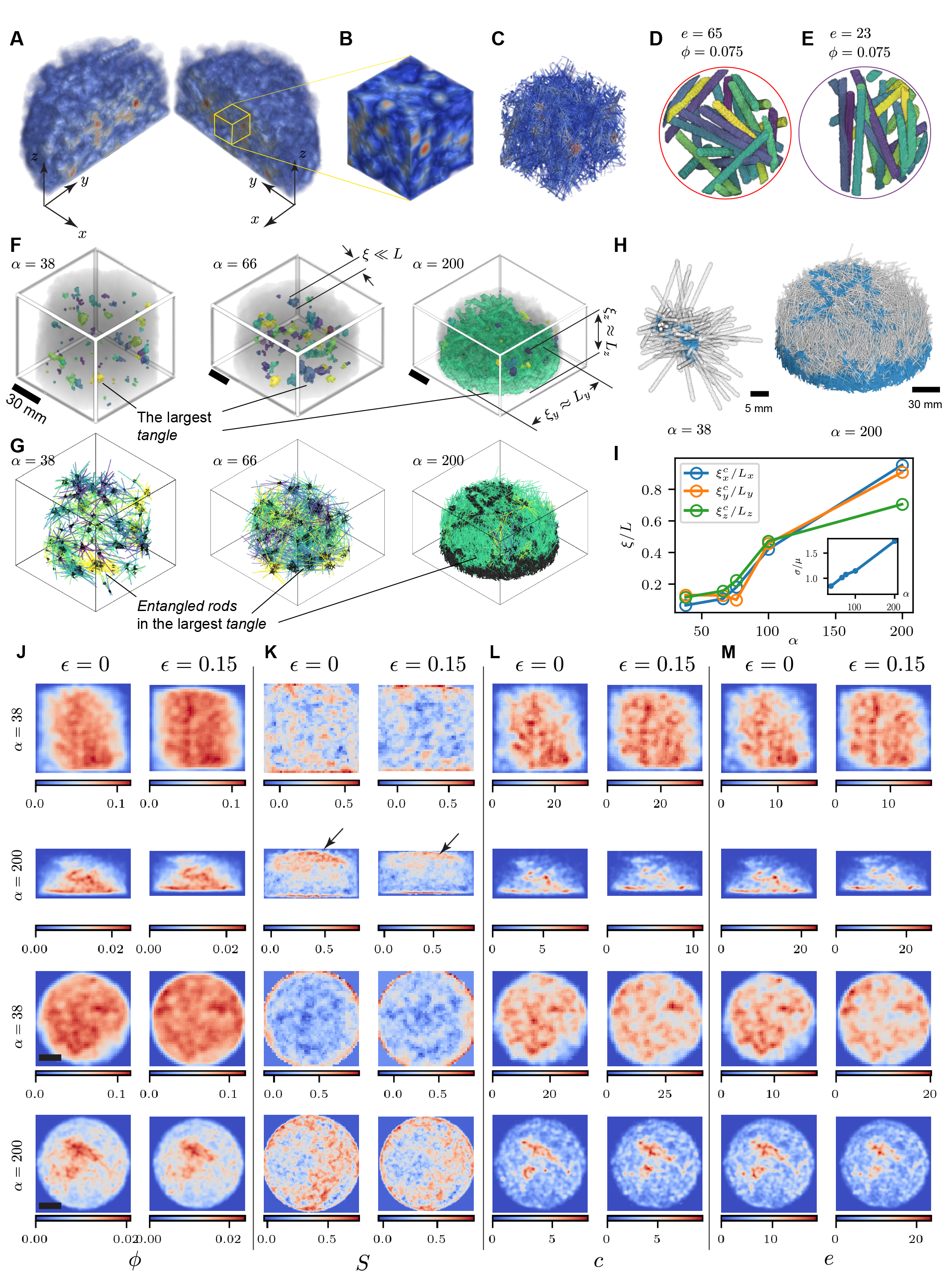}}
  \caption{\footnotesize{{{\bf Visualizations of 3D mesoscopic fields and highly entangled regions.} (A-C) 3D rendering of entanglement field for the packing of rods with $\alpha = 100$. (A) Vertical cross-section image of entanglement field for $\alpha = 100$. (B) A zoomed-in view on local cube marked in the yellow box in (A). (C) 3D rendering of rods found in the local domain in (A) and (B) overlaid with the entanglement field. (D-E) Two distinct configuration of rods with different entanglement values, $e = 130$ for (D), and $e = 45$ for (E) for same volume fraction, $\phi = 0.075$. (F) The regions of outliers in $p(e)$ ($\mathbf{x}$ such that $e(\mathbf{x}) > e_\mathrm{lb}$ as defined in Fig.~\ref{fig:Fields}E and I) for different $\alpha \in \{38, 66, 200\}$. (G) \textit{Entangled} rods in \textit{tangles} of rod packings with $\alpha \in \{38,66,200\}$. Rods in tangles are grouped and colored identically to their corresponding blobs in (F). The tangles are colored black. (H) Visualizations of entangled rods in the largest tangles for $\alpha \in \{38,200\}$. Entangled rods and tangle regions are colored gray and blue, respectively. (I) The tangle dimensions, $\xi_x$, $\xi_y$, and $\xi_z$, as a function of $\alpha$. Inset: The cofficient of variation ($\sigma_e/\mu_e$) of entanglement field as a function of $\alpha$. (J-M) Visualizations of horizontal-averaging (first 2 rows) and vertical-averaging (last two rows) (see (A)) of $\phi$ (J), $S$ (K), $c$ (L), and $e$ (M) fields.  The results are shown for two different strains, $\epsilon = 0$ and $\epsilon = 0.15$.}}}
  \label{fig:Visual}
\end{figure*}

Once we have the volume fraction, orientational order, contact number and the entanglement, we can evaluate the spatially integrated probability densities $p(\phi)$, $p(S)$, $p(c)$, and $p(e)$ for packings of rods for two extreme values of $\alpha = 38$, $\alpha = 200$ are shown in Fig.~\ref{fig:Fields}B-E and Fig.~\ref{fig:Fields}F-I, respectively, for different values of applied mechanical strain $\epsilon \in [0,0.15]$. The spatial distributions of these mesoscopic fields are shown in Fig.~\ref{fig:Visual}J-M for the same values of $\alpha$ and $\epsilon$. In general, the spatial average, denoted as $\langle (\cdot) \rangle = (1/V) \int_V (\cdot) \mathrm{d}^3\mathbf{x}$, of a local quantity $(\cdot)$ does not straightforwardly correspond to the equivalent global quantity obtained as $R \rightarrow V^{1/3}$ with an arbitrary choice of $R$. However, with $R = (dl^2)$, the spatial average of local volume fraction, $\langle \phi \rangle$, shows the agreement with the global volume fraction $\Phi$ because $\langle \phi \rangle \sim \langle n \rangle d^2 R / V_\Omega \sim Z /\alpha$.

The spatial average of the local volume fraction, $\langle \phi \rangle \approx$, also follows the same relation, $\langle \phi \rangle \sim Z/\alpha$ as Eq.~\ref{eq:volume_fraction}. Indeed, the shape of the probability distribution of $\phi$ is supposed to be independent of $\alpha$ since $p(\phi) \approx p(n) d^2R/R^3 \sim p(n) \alpha$, or the probability distribution function of higher $\alpha$ is simply a scaled version of that for low $\alpha$, as long as the orientational statistics of a packing remains random and isotropic. However, we see that there is a qualitative change in $p(\phi)$ with respect to $\alpha$ shown in Fig.~\ref{fig:Fields}B and F. This indicates that there is a qualitative change in orientational statistics between $\alpha = 38$ and $\alpha = 200$. The distribution $p(S)$ of orientational order parameter $S$ for these different values of $\alpha$ confirms that the packing of higher $\alpha$ becomes more aligned ($\langle S \rangle \gg 0$) as shown in Fig.~\ref{fig:Fields}C and G.

The contact density $p(c)$ and entanglement distribution $p(e)$ also show a similar transition as a function of $\alpha$.  Compared to the distributions for shorter rods ($\alpha = 38$), long slender rod packing ($\alpha = 200$), $p(c)$ and $p(e)$ show long-tailed distributions, although $p(e)$ exhibits a longer and fatter tail than $p(c)$ does. 

This suggests that the entanglement $e$ is more sensitive than contact $c$ to variations in the positional and orientational order of the rods. Indeed Fig.~\ref{fig:Visual}A-E (see also Video S4) allows us to visualize the 3D structure of the entanglements on small and intermediate scales. To probe the entangled regions more quantitatively, in Fig.~\ref{fig:Visual}F we display the spatial location of the outlier clusters in $e(\mathbf{x})$, defined by regions where $e > e_\mathrm{lb} = Q_1 + 1.5(Q_3-Q_1)$ with $Q_1$ and $Q_3$ being the lower and upper quartiles.   For $\alpha \ll 100$, these clusters are sparse and disconnected, but for $\alpha \geq 100$, the size of these clusters increases with $\alpha$ and eventually percolates through the system. We observe the percolation behavior independent of the choice of outlier detection methods (see \textbf{Supplementary Text} 3: Outlier analysis of entanglement field, and Fig.~S8). In Fig.~\ref{fig:Visual}G,  we show the location of these clustered \textit{tangles}, which increase in size with $\alpha$. The entangled rods in the largest \textit{tangles} are also shown in Fig.~\ref{fig:Visual}H (see Fig.~S7 for visualizations of tangles for packings with $\alpha \in \{38,66,76,100,200\}$). Denoting the dimensions of a tangle by $\xi_x$, $\xi_y$, and $\xi_z$, we see that rods with small aspect ratio form small tangles, i.e. $\xi_i \ll L_i$ (where $i$ labels $x$, $y$, or $z$), while those of rods with large aspect ratio form tangles with dimensions comparable to the system size. In Fig.\ref{fig:Visual}I, we see that the ratio $\xi_i/L_i$ is a function of the aspect ratio $\alpha$ and exhibits an abrupt transition as $\alpha \ge 100$ mirroring the change in mechanical response observed in Fig.~\ref{fig:Intro}F.

To characterize the heterogeneities of the 3D fields shown in Fig.~\ref{fig:Visual}A we average over the horizontal ($x$) and vertical ($z$) directions. In Figs.~\ref{fig:Visual}J-M, the packings of rods with $\alpha = 38$ exhibit more uniform spatial distributions in the fields $\phi(\mathbf{x}), S(\mathbf{x}), c(\mathbf{x}),~e(\mathbf{x})$ (with short-range fluctuations). For rods with $\alpha = 200$, we see the appearance of much more strongly heterogeneous variations in the mesoscopic geometric fields. For example, the orientational order $S$ shows the presence of a band of aligned rods, $S > 0.5$ (see the arrows in Fig.~\ref{fig:Visual}K), a feature not seen for rods with $\alpha = 38$. The hot spots of $c$ and $e$ shown in Fig.~\ref{fig:Visual}L and M increase in number and relative intensity in response to mechanical strain $\epsilon$, consistent with an enhanced mechanical and dynamical stability described in Fig.~\ref{fig:Intro} (see Section~S1, Fig.~S9-S12, which shows the gradual change of the spatially varying fields, $\phi(\mathbf{x})$, $S(\mathbf{x})$, $c(\mathbf{x})$, and $e(\mathbf{x})$ for $\alpha = \{38, 66, 76, 100, 200\}$).

A useful analysis of the heterogeneities within the entanglement field is afforded by the first two moments and the coefficient of variation of the statistics of entanglement, $\sigma_e/\mu_e = \frac{\sqrt{\langle e^2 \rangle - \langle e \rangle^2}}{\langle e \rangle}$ as shown in Fig.~S13. We observe that the mean does not correlate well with mechanical stability, largely because the choice of $R_\Omega$ was based on a simple geometrical model. Instead, the coefficient of variation $\sigma_e/\mu_e$ is a monotonically increasing function of $\alpha$, demonstrating a strong correlation with mechanical stability. The use of the coefficient of variation of the entanglement allows us to consider two-body interactions while avoiding the effects of sampling (by normalizing using the mean), and so from now on we will use $\sigma_e/\mu_e$ as an order parameter to define a phase diagram for the entanglement transition in rod packings, and show that this parameter approaches unity at the transition.


Since the rod packings are created in and subject to gravity, we might expect to see a stratification induced by this global field. Indeed, upon coarse-graining over the horizontal dimensions  we find that the fields $\phi(z)$, $S(z)$, $c(z)$, and $e(z)$ (as defined in \textbf{Supplementary Text 5}) exhibit distinctly different behaviors as a function of height, with a transition in the entanglement in the neighborhood of $z \approx l/2$, ostensibly due to the interplay between gravitational forces  and the boundary effects, observed in both experimental (Fig.~S14) and numerical packings (Fig.\ref{fig:sim}G).

\subsection*{Numerical simulations of rod packings}

To get a quantitative comparative baseline for the physical rod packings that also allows us to move through a larger range of parameter values than experimentally possible, we use numerical simulations, which proceed in two stages: (i) we first generate a packing of randomly oriented rods that do not intersect with each other or with the container (see \textbf{Materials and Methods} \ref{sec:numerical-simulation}), and (ii) we then simulate the packing of rods under mechanical excitation using an open-source package (`disMech') \cite{choi_dismech_2024}, customized to accelerate computation. We refer to these stages as \textit{Random Packing} and \textit{Entangled Packing}, respectively. In stage (i), we tentatively add randomly positioned and oriented rods in the container that are not intersecting with the container wall and the previously added rods. We repeat this addition until the number of rods in the containers reaches the desired value calculated from (\ref{eq:volume_fraction}), corresponding to a given density. In stage (ii), we use the `disMech' package which adopts an implicit contact model \cite{choi_implicit_2021,tong_fully_2023} to characterize rod-rod contacts. (See Videos S5 and S6 for the evolution of the Entangled Packing under mechanical excitation from the Random Packing and \ref{sec:numerical-simulation} for detailed descriptions.).

Random packings (RPs) use rods of varying aspect ratios $\alpha \in (25,300)$ contained within a box of dimensions ($1 \times 1 \times 2.5$ rod length). We fixed the rod length and varied the rod diameter to control the aspect ratio. Each random packing configuration contains a different number of rods within the same container volume to achieve a consistent number of contacts per rod following Eq.~\ref{eq:volume_fraction}. 
We then evolve the random packings dynamically to create entangled packings (EPs) via mechanical excitation with an externally imposed acceleration $a$ in a gravitational field $g$. In our simulations, we assume that the scaled acceleration $a/g \in [10^{-3},10^3]$ by choosing $A \approx 0.001l$ and $f \in [1,1000]$~Hz, and the scaled static load ( defined in terms of the weight of a single rod) $F/(\rho_s g d^2 l) \in [0.5,20]$. The simulations used a fixed friction coefficient $\mu = 0.2$. We note that without friction rod packings are not stable, as shown in Video S10.

In Fig.~\ref{fig:sim}C, we show that the average number of contacts per rod increases rapidly in the initial stage and then levels off. Similarly, the coefficient of variation for the entanglement $\sigma_e/\mu_e$ rise quickly at first and then stabilizes. The distribution of entanglement, $e$, denoted as $p(e)$, in EPs exhibits a power-law behavior whereas the distribution of $c$, $p(c)$, does not adhere to a power-law, as illustrated in Fig.~\ref{fig:sim}E and F. Interestingly, these results deviate from the experimental observations presented in Fig.~\ref{fig:Visual}. We hypothesize that boundary effects arising from the numerical EPs being confined within a single-length-sized box may have influenced the observed statistical outcomes. Nevertheless, the clustering and percolation behavior with respect to $\alpha$ is consistent with the experimental observations as shown in Fig.~\ref{fig:sim}G and Fig.~S7F.

To assess the dynamic stability of EPs and compare them to our observations,  we apply a harmonic excitation to the system without any lateral confining boundaries. Like in experiments, packings made of large aspect-ratio rods form large stable clusters, while those with shorter rods fall apart. To characterize the stability of the packings, we define the ratio of the number of rods in the largest contact cluster to total number of rods in the packing as $f$, and see how this varies with the scaled acceleration parameter $a/g$, once this parameter becomes relatively constant (noting that there is likely to be a slow aging as in granular packings \cite{ono_effective_2002,chakraborty_jamming_2009}, which we ignore here).  In Fig.~\ref{fig:sim}H, we show that $f$ increases with the aspect ratio $\alpha$; indeed the dynamical behavior of small and large aspect ratio packings are qualitatively differently under mechanical excitation, as shown in Videos S5 and S6. We can define the critical aspect ratio $\alpha^*$ separating the stable entangled and unstable untangled phases in terms of the characteristic excitation acceleration $a = 4\pi^2 f^2 A$ and static loading $\tilde{F} = F/\rho_s g d^2 l$. We find that for $g \approx 10$ m/s$^2$ and and $a/g \approx 1$, the critical parameter $\alpha^*\sim 100$ for both experiments and simulations. Furthermore, we see that the dependence of $\alpha^*$ on the scaled dynamic loading $a/g$ is less sensitive compared to that on the static loading $\tilde{F} = F/\rho_s g d^2 l = g_\mathrm{eff}/g$ - Fig.~\ref{fig:sim}I. Overall, we find that the mechanical stability to static and dynamic loadings observed in experiments aligns with our numerical simulations qualitatively; in both cases for entangled packings, $\alpha^* \approx 100$ and $\sigma_e/\mu_e \approx 1$  for both dynamic and static loadings characterized by $F/(\rho_s g d^2 l) \approx 1$ and $a/g = 2.5$, respectively. All together, our results suggest that topological constraints that arise with an increase in $\alpha$ significantly influence the formation of \textit{tangles} in real and numerical rod packings, as depicted in Fig.~\ref{fig:Visual}F-H and Fig.~S7. 

\subsection*{Entanglement phase diagram}
\begin{figure*}
  \centering
  \includegraphics[scale = 0.7]{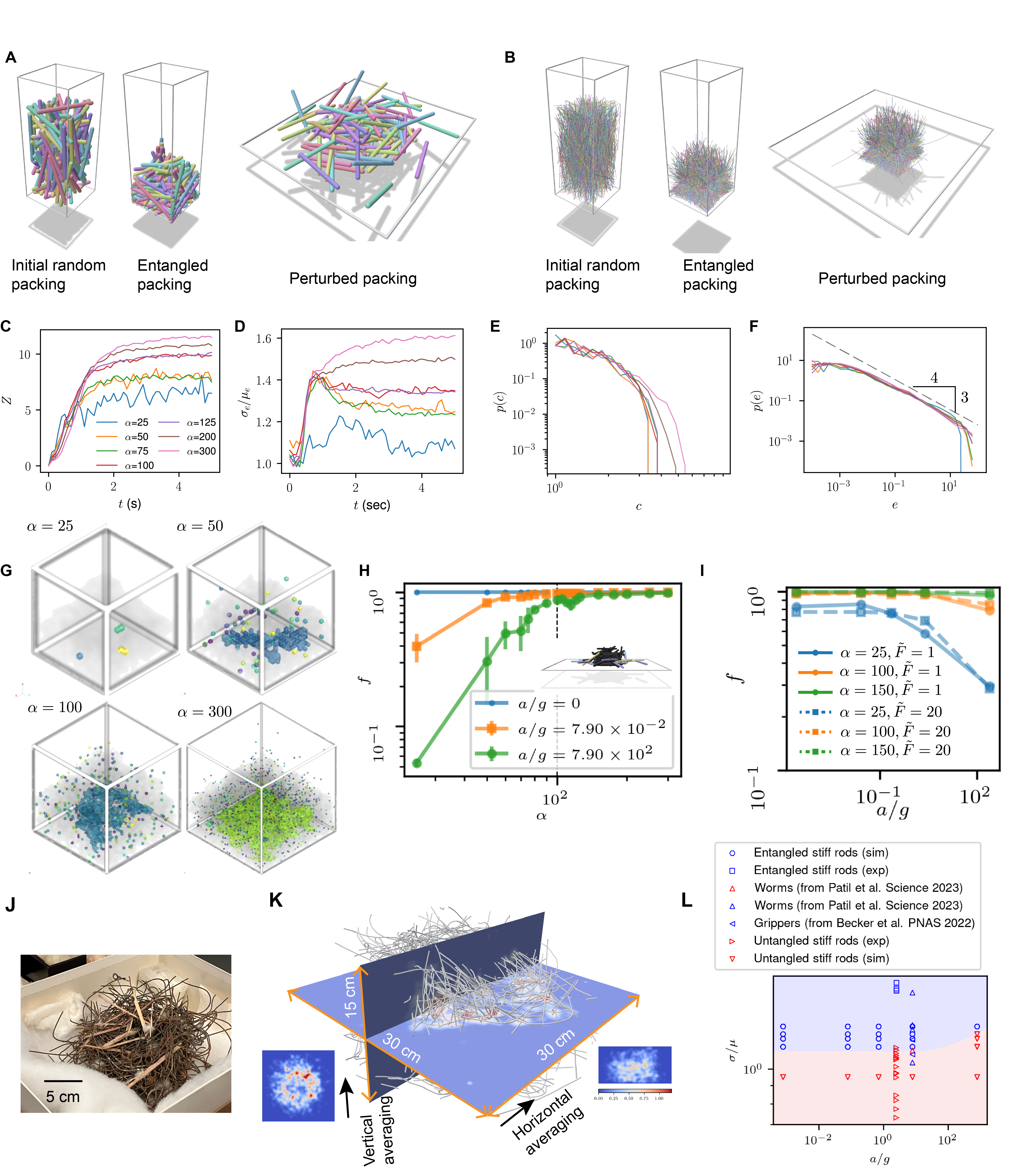}
  \caption{\textbf{Entanglement phase transition.} (A and B) Numerically generated Random Packings (RPs), their transformed Entangled Packings (EPs), and perturbation results by mechanical excitation ($\omega = 10$~Hz and $A = 0.0025$~mm) for $\alpha = 25$ (A) and $\alpha = 300$ (B). (C) The evolution of average number of contacts per rod in numerical rod packings of $\alpha = 25 - 300$ during the mechanical excitation ($\omega = 10$ Hz, $A = 0.0025$ mm). (D) The evolution of coefficients of variability ($\sigma_e/\mu_e$) for the entanglement field. (E and F) Probability density functions of $e$ (E) and $c$ (F) for $\alpha = 25 - 300$ in loglog scales. (G) The regions of outliers in $p(e)$ ($\mathbf{x}$ such that $e(\mathbf{x}) > e_\mathrm{lb}$ for different $\alpha \in \{25, 50, 100, 200\}$. (H) Fraction of the largest contact cluster, $f$, as a function of the aspect ratio $\alpha$ for different dynamic loads ($a/g = 0$, $7.90\times 10^{-2}$, and $7.90\times 10^{2}$) where $a = 4\pi A^2 \omega^2$. (I) Fraction of the largest contact cluster, $f$, as a function of dynamic loads ($a/g$) with different $\alpha$ and static loads $\tilde{F} = F/\rho_s g d^2 l$. (J) A nest of Pigeon (\textit{C. livia}) made of metallic wires\cite{MCZ_Orn_364335} (K) Vertical and horizontal cross sections of entanglement fields at the mid planes of the nest shown in (J). Filaments are also displayed for a visual aid. (L) The entanglement phase diagram is presented with respect to dynamic load ($a/g$) and the coefficient of variation ($\sigma_e/\mu_e$), while the static load is kept constant ($\tilde{F} = F/\rho_\mathrm{s} g d^2 l = 1$). The blue and red shades in the figure are included as visual aids.}
  \label{fig:sim}
\end{figure*}

Our experimental and computational studies on rod packings show that the aspect ratio controls the mechanical response of the  structure via changes in the geometry and topology of the network, as characterized especially by the inhomogeneous distribution of entanglement field $e$. As the coefficient of variation $\sigma_e/\mu_e$ increases (due to entangling process for high aspect ratio packings), we observe a transition between a stable entangled phase and an untangled phase (blue and red regions in Fig.~\ref{fig:sim}L, respectively), suggesting its use as a natural order parameter. This suggests a phase diagram  shown in Fig.~\ref{fig:sim}L that characterizes the regime for stable entanglement of rods in terms of three natural variables, the coefficient of variation of the entanglement field, $\sigma_e/\mu_e$, in the spatial distribution of the field $e$, a scaled measure of dynamic loading $a/g$, and a scaled measure of static loading $F/\rho_s g d^2 l$ (see also Fig.~S16). While there is a similarity to the jamming phase diagram for localized particles \cite{liu_2014}, where the temperature, applied stress and reciprocal of volume fraction are used to characterize the jamming transition, via a superficial analogy between the dynamic load and temperature and static load and stress, our statistical measure of the entanglement in terms of $\sigma_e/\mu_e$ is qualitatively different and reflects the extended nature of rods that we believe requires a measure that is non-local. We note that the static load does not significantly affect stability, as shown in Fig.~\ref{fig:sim}I. However, we can enhance the phase diagram by incorporating all three axes to create a 3D version, as illustrated in Fig.~S16, where we see that along the $F/\rho_s g d^2 l$ axis, there is minimal change in packing stability.

To test the effectiveness of our phase diagram, we go beyond our experiments and simulations by first considering an natural functional example of rod packing, the nest of pigeon (\textit{C. livia}) \cite{MCZ_Orn_364335}, made of metallic wires gathered from a construction site (Fig.~\ref{fig:sim}J). The entanglement field of the nest is shown in Fig.~\ref{fig:sim}K and Video S7 and is seen to be highly heterogeneous, with regions of high entanglement forming a network of entangled wires. From the field, we obtained the order parameter for entanglement, $\sigma_e/\mu_e \approx 1.4$., and falls within the regime corresponding to a stable entangled phase in Fig.~\ref{fig:sim}L. We then retrieved data\cite{becker_data_2022,patil_data_2023} from previous studies on topological robotic grasping \cite{becker_active_2022} and active switching between entangling/untangling states in worms \cite{patil2022ultrafast} and confirmed that their coefficient of variation is also much larger than unity (see Fig.~\ref{fig:sim}L and Fig.~S17). Finally, we utilized the available geometry datasets to create numerical simulations of the systems, assuming the individual filaments are naturally curved as in the scanned state, under a similar scaled excitation acceleration of $a/g \approx 10$ as we did for most of our experimental and numerical rod packings. (see Videos S8 and S9). We see that all these packings are dynamically stable. All together, these results suggest that the entanglement phase diagram can be a useful tool for understanding the mechanical stability of entangled systems in natural and artificial settings.

\section*{Discussion}

Inspired by simple observations of the robust mechanical stability of packings of long stiff rods, we used a combination of X-ray CT imaging and numerical simulations of stiff frictionally contacting rods to investigate and quantify a number of different mesoscopic geometric fields that characterize the geometry and topology of these systems. Our results show the emergence of a stable packing that arises via the \textit{percolation of tangles, zones of high entanglement,} when the rod aspect ratio is higher than a threshold ($\approx 100$), linking mechanical rigidity and the statistics of a mesoscopic non-local object, the entanglement field. We interpret this in terms of the coefficient of variation $\sigma_e/\mu_e$ of the entanglement probability distribution $p(e)$ which increases monotonically with the aspect ratio in the physical packings, contrary to what is known in random packings \cite{Philipse1996}. Using this quantity as an order parameter allows us to define an entanglement transition wherein the packing becomes stiff when the entanglement order parameter, $\sigma_e/\mu_e\approx1.4$, becomes larger than unity for a packing of rods with $\alpha^* = 100$ for $a/g \approx 1$ and $F/(\rho_s g d^2 l) \approx 1$. We confirm this using our experimental data along with a reanalysis of published data from different systems that range from bird nests to robotic graspers and worm blobs \cite{becker_data_2022,patil_data_2023}. Our numerical simulations of stiff rods are also consistent with the experimental observations in many aspects, including the formation and percolation of highly entangled regions (Fig.~\ref{fig:sim}G) and the transition behavior near $\sigma_e/\mu_e \approx 1.4$ (Fig.~\ref{fig:sim}D and L).


We close by emphasizing that the role of entanglement outliers and heterogeneities in packings of stiff rods is also well-defined for packings of flexible filaments as well, as we have deployed them to bird nests, worm swarms  and robotic systems. As the entanglement order parameter $\sigma_e/\mu_e \approx 1.4$ exceeds unity, both experiments and numerical simulations show the appearance of mechanically stable packings. As we continue to think about the development of filamentous architectures\cite{dierichs_towards_2016,bonavia_minimal_2023}, active fabrics\cite{raviv2014active}, and soft robots \cite{becker_active_2022}, this simple  principle which embodies the geometrical and topological interactions between constituents might serve as a way to characterize and distinguish successful designs and active approaches for reconfigurable packings. Understanding how the critical value of this order parameter depends on the stiffness of the rods and the physics of their interaction through friction, lubrication, etc., awaits further study, which we hope will be inspired by the preliminary steps outlined here.

\section*{Acknowledgments}

We appreciate the Museum of Comparative Zoology at Harvard University for lending the specimen (MCZ Ornithology 364335) and assisting with the x-ray scanning.

{\bf Funding.} This research was partially supported by National Research Foundation of Korea\\
(Grant No. 2021R1A6A3A03039239), and the Harvard NSF MRSEC 20-11754, NSF DMREF 19-22321, the Simons Foundation and the Henri Seydoux Fund, and performed in part at the Harvard University Center for Nanoscale Systems (CNS); a member of the National Nanotechnology Coordinated Infrastructure Network (NNCI), which is supported by the National Science Foundation under NSF award no. ECCS-2025158. 

{\bf Author contributions.} L.M. conceived of the question and approaches. Y.J., T.P-R. and L.M. designed the study. Y.J. and T.P-R. carried out the experiments with the help of G.L. Y.J. and L.M. analyzed the data. Y.J. and L.M. wrote the paper.  L.M. supervised the study. The authors declare no conflict of interest.

{\bf Competing interests.} The authors declare no competing interests.

{\bf Data availability.} All data are available in the main text or the supplementary materials.

\section*{Supplementary materials}
Figs. S1 to S17\\
Table S1 to S4\\
Sections S1 to S7\\
Videos S1 to S10\\

\section*{Materials and Methods}
\subsection*{General consideration in rod packing preparation}
\label{sec:packing-preparation}

We review the theory of random rod packing as outlined by \cite{Philipse1996} to validate our selection of parameters in packing protocols. The excluded volume between two cylinders, each with diameter \(d\) and length \(l\) (as per \cite{Onsager1949}), is expressed as:
\[
    V_{\text{ex}} \sim d l^2 |\mathbf{u}_i \times \mathbf{u}_j|
\]
This calculation considers only edge-to-edge collisions, neglecting other types of interactions, which is particularly relevant for high aspect ratio packings. In a given system volume where the centroid of a rod can be freely positioned, the probability of contact is given by:
\[
    P = \frac{V_{\text{ex}}}{V} = \frac{dl^2 |\mathbf{u}_i \times \mathbf{u}_j|}{V}
\]
By summing pairwise contributions over all possible pairs between $N$ rods within the volume \(V\), the expected total number of contacts is derived as:
\[
    C \sim \sum_{i,j} \frac{V_{\text{ex},ij}}{V} = \sum_{i,j} \frac{dl^2 |\mathbf{u}_i \times \mathbf{u}_j|}{V} \sim \frac{N^2 dl^2}{V}
\]
Thus, the expected number of contacts per rod is:
\[
    Z = \frac{C}{N} \sim \frac{N dl^2}{V}
\]
The number of contacts must satisfy the isostatic condition. Consequently, the average coordination number $Z$ must be consistent (or minimally variable) across different aspect ratios. Therefore, the number of rods \(N\) is determined by the rod dimensions \(d\) and \(l\), according to the relation:
\[
    N \sim \frac{Z V}{dl^2}
\]
By dividing this equation by the volume of the container, we obtain the number density of rods \(\rho\) in the container:
\[
    \rho = \frac{N}{V} \sim \frac{Z}{dl^2}
\]
Subsequently, the volume fraction \(\phi\) is given by:
\[
    \Phi = \frac{\pi N d^2 l}{V} \sim \frac{Z}{\alpha}
\]
where \(\alpha = d/l\) as shown in \cite{Philipse1996}.

\subsection*{Sample Preparation}
\label{sec:sample-preparation}
The constituent parts of the random packings were made by cutting stainless steel rods of varying diameters and length (see Table.~S4). Rods of the same dimensions were dropped steadily through a sieve and in small numbers—so as to avoid adding any pre-entangled groups—into a cylindrical container with a diameter at least three times the length of a single rod (see Table.~S4). This larger size minimized the undue influence of the container itself on the final orientation of the rods. In test runs while designing the experiment, we observed that as the container diameter approached the length of a single rod, the orientation of the rod network was strongly vertical. The rods were added to a height of $h\approx100$ mm, but the final height can depend on how rods pack in response to external excitation, which is predicted by the random contact model\cite{Philipse1996} and deviates from the prediction as the aspect ratio increases from $\alpha \approx 100$.

During the packing preparation, the container was jostled by a motor and inverted by hand several times to randomize rod orientations and reach a predicted steady-state (total) volume fraction $\Phi$ determined by the rod aspect ratio ($\Phi \alpha = Z$) with $Z$ being the critical number of contacts per rod. In preparation, we used $Z$ = 5.4 in the sample preparation and the experimental values with x-ray images was obtained as, $Z_\mathrm{exp} = 6.2 \pm 1.1$.

\subsection*{Mechanical Testing}
\label{sec:mechanical-testing}
In the excitation experiment, we placed approximately $N \approx 1.25\times10^6/(d l^2)$ steel rods into a cylindrical container with a diameter of 150 mm and a height of 150 mm. This quantity was chosen to ensure that the rod packing maintained a consistent total volume, with an average of approximately 6 contacts per rod, and exhibited random orientations. Initially, the rods were deposited into an inner tube to temporarily contain them. This tube was subsequently removed prior to the excitation phase. The excitation was performed at a frequency of 5 Hz with an amplitude of 25 mm (see Video S2).

\subsection*{X-ray Computerized Tomography imaging}
\label{sec:x-ray-ct}
X-ray Computerized Tomography (X-ray CT) imaging was completed on the X-Tek HMXST 225 system (Nikon Metrology, Inc., Brighton, Michigan, USA) at Harvard University Center for Nanoscale Systems.  Multiple diameters of steel rods forming network of different aspect ratios are packed inside each individual clear plastic cylindrical container of 150 mm diameter.  X-ray CT imaging was performed for each set of rods in the unstrained state, and also after applying each additional load and compressive strain.  Orientation of each plastic container was carefully tracked to observe how the rods shift with strain.  X-ray CT scanning was performed using a microfocus cone beam reflection tungsten target X-ray source with tube voltage of 150 kV and target current of 380 µA (57 W).  A 1.0 mm thick copper filter was implemented for all imaging to reduce the polychromaticity of the beam and remove beam hardening artifact.  3142 projection radiograph images were captured per sample on a 16-bits 2000 pixels by 2000 pixels flat panel detector with each exposure time at 500 ms.  The achievable voxel resolution for all scanning is around 78.22 $\mu$m.  Data was then reconstructed using CT Pro 3D software (Nikon Metrology, Inc., Brighton, Michigan, USA) and exported using VGStudio MAX software (Volume Graphics, Inc., Charlotte, North Carolina, USA) to 16-bits unsigned TIFF format for further image analysis.

\subsection*{Image Segmentation}
\label{sec:image-segmentation}
The MATLAB codes used for image segmentation in this study are available on Zenodo \cite{jung_2024_13382297}. Below, specific custom MATLAB function names from the repository are explicitly referred to in the corresponding paragraphs.

The X-ray tomography images were initially filtered and binarized to distinguish foreground (rods) from background (void). Traditional methods, such as the watershed algorithm or skeletonization based on distance transform, often suffer from oversegmentation and branching issues, particularly because these methods are typically effective only for structures like barely touching ellipsoids, which do not represent the complexity of real images. This oversegmentation problem can be mitigated by the paste method \cite{Gomez2020}, which merges oversegmented points. However, we found that the paste method becomes less effective when dealing with a large number of contacts, as point clouds near contacts may be completely lost. The Local Template Matching method \cite{rigort_automated_2012} can accurately locate the centerlines of filamentous structures without oversegmentation issues, though it usually incurs a high computational cost.

The image processing algorithm used in this study combines skeletonization and template matching methods, modified to accelerate the segmentation process. Instead of searching for a matched template across all foreground voxels, we start with a good initial guess by skeletonization (using the function \texttt{segmentation\textunderscore by\textunderscore bwskel}) to identify a centerline position where the adjacent point cloud matches a predefined cylinder with a high goodness-of-fit (see Video S3).

Once an initial point and corresponding cylinder are identified, it is likely that another matching cylinder will be found adjacent to the initially matched one. The initial point then progresses along the cylinder axis, fitting a cylinder to the nearby point cloud at the next point. This local cylinder matching process continues until the moving point reaches the background, at which point the process restarts at the initial point in the opposite direction. These successive local cylinder matching processes, implemented in \texttt{lengthen\textunderscore centerlines\textunderscore from\textunderscore ends}, generate reliable segments that are eventually 'pasted' together. To paste segments in \texttt{trim\textunderscore centerlines}, an optimization scheme is used where 1) the straightness of pasted segments, 2) the proximity of pasted segments, and 3) the length of each pasted segment are optimized rather than relying on conditional statements (e.g., ensuring that the pasted curve has a curvature below a specific threshold, or that the pasted length is shorter than the physical rod length) (see Fig.~S2).

Cylinder matching for a point cloud can be approached in various ways. In this study, we implemented a function that counts the number of voxels within a cylinder defined by its centroid, radius, and axis direction. Since a cylinder with a fixed length is defined by its centroid (three parameters) and orientation (two parameters), the function takes these five-dimensional inputs and outputs the number of points within the cylindrical boundary for a given point cloud. The argmax of this function was found using MATLAB's \texttt{fminsearch}, which employs the Nelder-Mead simplex algorithm \cite{Lagarias1998}.

In the final segmentation process, segments that are not reconnected through the successive cylinder matching process are reconnected using effective distance functions and clustering based on these distance functions. To extract contact statistics using the \texttt{get\textunderscore contact\textunderscore info} function, we first identified all rod pairs whose distance was less than a specified threshold (e.g., twice the rod diameter). We then confirmed that the binarized volume images of the rods were indeed in contact by identifying foreground voxels corresponding to each rod pair.

The accuracy of the segmentation was evaluated by generating rod packings with known positions, orientations, and contact points, and then comparing the segmentation results to this predefined data (see Fig.~S3 and \textbf{Supplementary Text 7}). This validation demonstrates that our rod segmentation technique can identify the rod centerlines and contact points within a margin of 1 pixel.

\subsection*{Percolation analysis}
\label{sec:percolation-analysis}
The values of $e(\mathbf{x})$ for all $\mathbf{x}$ are illustrated in the histograms of Fig.~\ref{fig:Fields}E and 2\ref{fig:Fields}I. To identify outliers in such non-Gaussian distributions, we use the lower and upper quartile values, $Q_1$ and $Q_3$. We define the lower bound for entanglement outliers, $e_\mathrm{lb}$, as $e_\mathrm{lb} = Q_1 + 1.5(Q_3 - Q_1)$.

The entanglement field $e(\mathbf{x})$ is a continuous scalar function conceptually, but in our analysis, it is approximated by a three-dimensional discrete array. This analysis reveals clusters of grid points with entanglement values, $e(\mathbf{x})$, exceeding the lower outlier bound $e_\mathrm{lb}$, as depicted in Fig.~\ref{fig:Visual}F. These clusters, or \textit{tangles}, are regions of exceedingly high entanglement within the rod packing.

The volume of each tangle is quantified by the count of grid points comprising it. This count is normalized against the total number of grid points in the entanglement field, $e(\mathbf{x})$, representing the system's volume. Consequently, the relative tangle volume indicates the emergence of a substantial, interconnected tangle with an increase in $\alpha$, as shown in the far-right bin of the histogram in Fig.~S7D. The dimensions of a tangle, denoted as $\xi_x$, $\xi_y$, and $\xi_z$, correspond to the smallest box enclosing it.

\subsection*{Numerically generated rod packings}
\label{sec:numerical-simulation}

We set $l = 50~mm$ and varied $d$ to vary the aspect ratio $\alpha = d/l$. Before excitation, initially, the number of rods were set by the relation $\rho d l^2 = Z$ (from Eq.~\ref{eq:volume_fraction}) with $\rho = N/V$ and $V = l^3$, to aim for the same bounding volume for packings of different aspect ratios. The number of rods in the container is then derived to be $N = Z\alpha$. We set $Z$ to be 5.4 initially, but the actual coordination number along with the actual bounding volume would be determined by the dynamics of entanglement formation during the excitation.

The initial random packing of rods were prepared by a protocol that does not allow intersection between rods. We tentatively add a rod by generating two random vectors $\mathbf{x}_0, \mathbf{x}_1$ in a container $[0,L_x]\times[0,L_y]\times[0,L_z] \subset \mathbb{R}^3$ where $\mathbf{x}$ is three uniform random numbers and $\mathbf{x}_1 = \mathbf{x}_0 + l\mathbf{u}$ with $\mathbf{u}$ being a unit vector uniformly distributed on a sphere. Then, we check if the rod intersects with any of the previously added rods or the container wall. If the rod intersects with any of them, the rod is rejected. This process is repeated until the number of rods in the container reaches $N = Z\alpha$.

From these random packings as initial conditions, we numerically simulate the time evolution of rods under various scenarios (Videos S5 and S6). The simulations were carried out using an open source package for soft filaments \cite{choi_implicit_2021} which was customized in order to accelerate the computation, especilly the collision detection part. The parameters used in the simulation are summarized in Table \ref{tab:sim_params}. Rod aspect ratios ranging from 25 to 300 were selected, with the rod length fixed at 1 m, resulting in rod diameters varying from 1/25 to 1/300 m. The contact tolerance was set to 1/10 of the rod diameter. The time step for the Verlet numerical integration scheme was set to 10$^{-5}$ seconds.

For the simulation we used a scaled-up geometry rather than using experimental dimensions to avoid rounding errors, but kept the dimensionless parameters (relative gravity and aspect ratio) the same.  Specifically, we used the physical properties of steel rods but with 20 times larger geometry ($d \rightarrow 20 d$ and $l \rightarrow 20 l$) than what they really are. For the physical accuracy of the data, we also rescaled the gravitational acceleration to $g = 0.5 m/s^2$ so that the rescaled geometry is balanced with the rescaled gravitational accleration. In other words, to have the right relative balance between gravitational force and elastic bending force, we matched the torque exerted by contact forces to scale with the weight of rods and the internal bending reaction, or $(\rho d^2 l g)l \sim E I \kappa$. Given $\kappa \sim \delta / l^2$ with $\delta$ being the characteristic deflection, $g$ must satisfy $g = \frac{E I \delta}{\rho d^2 l^4} \sim \frac{E d^2 \delta}{\rho l^4}$. Thue fixing $E$ and $\rho$, $g$ should be scaled as $g \sim \delta d^2/l^4 g_0 \sim g_0/20$, with $g_0 = 9.8$~m/s$^2$, because $\delta \rightarrow 20 \delta$ too.

It is also noteworthy that the effect of gravitational acceleration is negligible, as demonstrated in Fig.~\ref{fig:sim}I. The other parameters were set to match the experimental conditions. Each rod is represented by a chain of 10 vertices (hence 9 edges) whose elastic potential energies were calculated following numerical implementations in 
\cite{choi_implicit_2021}. 

The initial random packings were first excited by a vertical vibration with a frequency of 10 Hz and an amplitude of 0.05 m. After the five seconds of entangling procedure, mechanical properties of entangled packings were assessed by dynamic excitation test with different frequencies and gravitational accelerations. The amplitude was fixed. In this test, the confinement was removed except the floor, and the entire system was vertically excited with a frequency $f \in [0.1, 1, 3, 10, 100]$  and an amplitude of 0.001 m with $g \in [0.5,2,10] m/s^2$ for 5 seconds.

\subsection*{Numerical simulation of x-ray scanned filaments for perturbation-induced stability}
\label{sec:numerical-simulation-perturbation}
We utilized digital data from previous studies \cite{becker_active_2022,patil2022ultrafast} to measure the entanglement field and numerically perturb the data to assess stability. We set \( a/g = 7.8 \), aligning with the majority of experimental and simulation data.
The parameters for each simulation were identical to those in the original papers if available, or we used known material properties otherwise. These parameters are summarized in Table S2, S3.  
We rescaled the filament data to ensure consistency with the parameters used in Entangled Packings. Specifically, the geometry of each filament was increased by a factor of 20. To ensure the physical accuracy of the dynamics, gravitational accelerations were rescaled to \( g = 0.5 \, \text{m/s}^2 \) as we did for Entangled Packings.

\clearpage

\clearpage

\clearpage

\clearpage

\clearpage


\begin{thebibliography}{10}

    \bibitem{ohern_random_2002}
    C.~S. O'Hern, S.~A. Langer, A.~J. Liu, S.~R. Nagel, {\it Physical Review
        Letters\/} {\bf 88}, 075507 (2002).
    
    \bibitem{ohern_jamming_2003}
    C.~S. O’Hern, L.~E. Silbert, A.~J. Liu, S.~R. Nagel, {\it Physical Review
        E\/} {\bf 68}, 011306 (2003).
    
    \bibitem{song_phase_2008}
    C.~Song, P.~Wang, H.~A. Makse, {\it Nature\/} {\bf 453}, 629 (2008).
    
    \bibitem{Liu2010}
    A.~J. Liu, S.~R. Nagel, {\it Annual Review of Condensed Matter Physics\/} {\bf
        1}, 347 (2010).
    
    \bibitem{liu_2014}
    A.~J. Liu, S.~R. Nagel, eds., {\it Jamming and rheology: Constrained dynamics
        on microscopic and macroscopic scales\/} (CRC Press, London, 2014).
    
    \bibitem{Baule2018}
    A.~Baule, F.~Morone, H.~J. Herrmann, H.~A. Makse, {\it Rev. Mod. Phys.\/} {\bf
        90}, 15006 (2018).
    
    \bibitem{Ekman2014}
    A.~Ekman, A.~Miettinen, T.~Tallinen, J.~Timonen, {\it Phys. Rev. Lett.\/} {\bf
        113}, 1 (2014).
    
    \bibitem{Philipse1996}
    A.~P. Philipse, {\it Langmuir\/} {\bf 12}, 5971 (1996).
    
    \bibitem{bi_jamming_2011}
    D.~Bi, J.~Zhang, B.~Chakraborty, R.~P. Behringer, {\it Nature\/} {\bf 480}, 355
        (2011). Number: 7377 Publisher: Nature Publishing Group.
    
    \bibitem{weiner_mechanics_2020}
    N.~Weiner, Y.~Bhosale, M.~Gazzola, H.~King, {\it Journal of Applied Physics\/}
        {\bf 127}, 050902 (2020).
    
    \bibitem{Bhosale2022}
    Y.~Bhosale, {\it et~al.\/}, {\it Phys. Rev. Lett.\/} {\bf 128}, 198003 (2022).
    
    \bibitem{Onsager1949}
    L.~Onsager, {\it Annals of the New York Academy of Sciences\/} {\bf 51}, 627
        (1949).
    
    \bibitem{Edwards1967}
    S.~F. Edwards, {\it Proceedings of the Physical Society\/} {\bf 91}, 513
        (1967).
    
    \bibitem{Doi1978}
    M.~Doi, S.~F. Edwards, {\it J. Chem. Soc. Faraday Trans. 2 Mol. Chem. Phys.\/}
        {\bf 74}, 560 (1978).
    
    \bibitem{hearle_physical_2008}
    J.~W.~S. Hearle, W.~E. Morton, {\it Physical Properties of Textile Fibres\/}
        (Elsevier Science, 2008).
    
    \bibitem{kabla_nonlinear_2007}
    A.~Kabla, L.~Mahadevan, {\it Journal of The Royal Society Interface\/} {\bf 4},
        99 (2007).
    
    \bibitem{buck_spectrum_2012}
    G.~Buck, J.~Simon, {\it Proceedings of the Royal Society A: Mathematical,
        Physical and Engineering Sciences\/} {\bf 468}, 4024 (2012).
    
    \bibitem{panagiotou_knot_2020}
    E.~Panagiotou, L.~H. Kauffman, {\it Proceedings of the Royal Society A:
        Mathematical, Physical and Engineering Sciences\/} {\bf 476}, 20200124
        (2020).
    
    \bibitem{patil_topological_2020}
    V.~P. Patil, J.~D. Sandt, M.~Kolle, J.~Dunkel, {\it Science\/} {\bf 367}, 71
        (2020).
    
    \bibitem{glover_measuring_2024}
    C.~Glover, A.-L. Barab\'asi, {\it Phys. Rev. Lett.\/} {\bf 133}, 077401 (2024).
    
    \bibitem{trepanier_column_2010}
    M.~Trepanier, S.~V. Franklin, {\it Physical Review E\/} {\bf 82}, 011308
        (2010). Publisher: American Physical Society.
    
    \bibitem{franklin_geometric_2012}
    S.~V. Franklin, {\it Physics Today\/} {\bf 65}, 70 (2012). Publisher: American
        Institute of Physics.
    
    \bibitem{heussinger_collapse_2023}
    C.~Heussinger, Collapse of columns of granular rods (2023). ArXiv:2301.05596
        [cond-mat].
    
    \bibitem{Gravish2012}
    N.~Gravish, S.~V. Franklin, D.~L. Hu, D.~I. Goldman, {\it Phys. Rev. Lett.\/}
        {\bf 108}, 208001 (2012).
    
    \bibitem{Zhao2016}
    Y.~Zhao, {\it et~al.\/}, {\it Granular Matter\/} {\bf 18} (2016).
    
    \bibitem{wang_structured_2021}
    Y.~Wang, L.~Li, D.~Hofmann, J.~E. Andrade, C.~Daraio, {\it Nature\/} {\bf 596},
        238 (2021).
    
    \bibitem{karapiperis_stress_2022}
    K.~Karapiperis, S.~Monfared, R.~B.~D. Macedo, S.~Richardson, J.~Andrade, {\it
        Granular Matter\/} {\bf 24}, 91 (2022).
    
    \bibitem{becker_active_2022}
    K.~Becker, {\it et~al.\/}, {\it Proceedings of the National Academy of
        Sciences\/} {\bf 119}, e2209819119 (2022).
    
    \bibitem{raviv2014active}
    D.~Raviv, {\it et~al.\/}, {\it Scientific reports\/} {\bf 4}, 7422 (2014).
    
    \bibitem{patil2022ultrafast}
    V.~P. Patil, {\it et~al.\/}, {\it Science\/} {\bf 380}, 392 (2023).
    
    \bibitem{bozdag_novo_2023}
    G.~O. Bozdag, {\it et~al.\/}, {\it Nature\/} {\bf 617}, 747 (2023).
    
    \bibitem{blouwolff_coordination_2006}
    J.~Blouwolff, S.~Fraden, {\it Europhysics Letters (EPL)\/} {\bf 76}, 1095
        (2006).
    
    \bibitem{choi_dismech_2024}
    A.~Choi, R.~Jing, A.~Sabelhaus, M.~K. Jawed, {\it IEEE Robotics and Automation
        Letters\/} {\bf 9}, 3483 (2024). ArXiv:2311.18126 [cs].
    
    \bibitem{choi_implicit_2021}
    A.~Choi, D.~Tong, M.~K. Jawed, J.~Joo, {\it Journal of Applied Mechanics\/}
        {\bf 88} (2021).
    
    \bibitem{tong_fully_2023}
    D.~Tong, A.~Choi, J.~Joo, M.~K. Jawed, {\it Extreme Mechanics Letters\/} {\bf
        58}, 101924 (2023).
    
    \bibitem{ono_effective_2002}
    I.~K. Ono, {\it et~al.\/}, {\it Physical Review Letters\/} {\bf 89}, 095703
        (2002).
    
    \bibitem{chakraborty_jamming_2009}
    B.~Chakraborty, B.~Behringer, {\it Jamming of granular matter\/} (Springer New
        York, New York, NY, 2009), pp. 4997--5021.
    
    \bibitem{MCZ_Orn_364335}
    M.~C.~Z. Harvard~University, Specimen mcz:orn:364335, {Amazona farinosa},
        cataloged by museum of comparative zoology, harvard university (2024).
        Accessed: 2024-08-12.
    
    \bibitem{becker_data_2022}
    K.~Becker, {\it et~al.\/}, harvard-microrobotics/entanglementgripper,
        {\it https://github.com/harvard-microrobotics/EntanglementGripper} (2022).
        Deposited 26 September 2022, GitHub.
    
    \bibitem{patil_data_2023}
    V.~P. Patil, {\it et~al.\/}, Datasets accompanying “ultrafast reversible
        self-assembly of living tangled matter”,
        {\it https://doi.org/10.5281/zenodo.7519337} (2023). Zenodo.
    
    \bibitem{dierichs_towards_2016}
    K.~Dierichs, A.~Menges, {\it Granular Matter\/} {\bf 18}, 25 (2016).
    
    \bibitem{bonavia_minimal_2023}
    E.~Bonavia, {\it et~al.\/}, {\it Construction Robotics\/} {\bf 7}, 329 (2023).
    
    \bibitem{jung_2024_13382297}
    Y.~Jung, T.~Plumb-Reyes, H.-Y.~G. Lin, L.~Mahadevan, {Replication Data for:
        Entanglement transition in random rod packings
        (https://doi.org/10.5281/zenodo.13382297)} (2024).
    
    \bibitem{Gomez2020}
    L.~R. G{\'{o}}mez, N.~A. Garc{\'{i}}a, T.~P{\"{o}}schel, {\it Proceedings of
        the National Academy of Sciences\/} {\bf 117}, 3382 (2020).
    
    \bibitem{rigort_automated_2012}
    A.~Rigort, {\it et~al.\/}, {\it Journal of Structural Biology\/} {\bf 177}, 135
        (2012).
    
    \bibitem{Lagarias1998}
    J.~C. Lagarias, J.~A. Reeds, M.~H. Wright, P.~E. Wright, {\it SIAM Journal on
        Optimization\/} {\bf 9}, 112 (1998).
    
    \end{thebibliography}
\end{document}